\documentclass[aps,pra,twocolumn,showpacs,superscriptaddress]{revtex4-1}
\usepackage[colorlinks ,linkcolor=blue,anchorcolor=blue,citecolor=blue,urlcolor=blue]{hyperref}
\usepackage{graphicx}
\usepackage{epstopdf}
\usepackage{soul, color, xcolor}
\usepackage{bm}
\usepackage{siunitx} 
\usepackage{amsmath,amssymb,amsfonts}
\usepackage{appendix}

\def\be{\begin{equation}} \def\ee{\end{equation}}
\def\bea{\begin{eqnarray}} \def\eea{\end{eqnarray}}

\begin{document}

\title{Magnetically controllable nonlinear valley Hall effect in centrosymmetric ferromagnets}

\author{Ruijing Fang}
\affiliation{School of Physics and Technology, Nanjing Normal University, Nanjing 210023, China}
\affiliation{Center for Quantum Transport and Thermal Energy Science,
Nanjing Normal University, Nanjing 210023, China} 

\author{Jie Zhang}
\affiliation{School of Physics and Technology, Nanjing Normal University, Nanjing 210023, China}
\affiliation{Center for Quantum Transport and Thermal Energy Science,
Nanjing Normal University, Nanjing 210023, China} 

\author{Zhichao Zhou}
\affiliation{School of Physics and Technology, Nanjing Normal University, Nanjing 210023, China}

\author{Xiao Li} 
\email{lixiao@njnu.edu.cn}
\affiliation{School of Physics and Technology, Nanjing Normal University, Nanjing 210023, China}
\affiliation{Center for Quantum Transport and Thermal Energy Science,
Nanjing Normal University, Nanjing 210023, China}

\begin{abstract} 
Valley Hall effect is fundamental to valleytronics and provides a promising avenue for advancing information technology.
While conventional valley Hall effect requires the inversion symmetry breaking, the recently proposed nonlinear valley Hall (NVH) effect removes the symmetry constraint, and broaden material choices. 
However, existing studies are limited to nonmagnetic materials without spin involvement and rely on external strain to break rotational symmetry.
Here, to address these limitations, we design a magnetically controllable NVH effect in centrosymmetric ferromagnets, by the tight-binding model and first-principles calculations.
The model calculations demonstrate nonvanishing NVH conductivities can emerge in pristine hexagonal lattice without external strain, with the magnitude, sign, and spin polarization of the conductivities being all dependent on the magnetization orientation.
The effect thus generates various spin-polarized valley Hall currents, characterized by distinct combinations of current direction and spin polarization.
First-principle results on a ferromagnetic VSi$_2$N$_4$ bilayer confirm considerable NVH conductivities and their dependence on the magnetization. 
The magnetically controllable NVH effect unlocks the potential of centrosymmetric magnets for valleytronics, and offer opportunities for novel spintronic and valleytronic devices.  
\end{abstract} 
\pacs{} 
\maketitle

{\color{blue}\textit{Introduction.}} -- 
Valleys, as energy extrema in electronic band structure, have become a rising degree of freedom with enormous applications in developing advanced electronic and optoelectronic devices \cite{Rycerz2007,Xiao2007,Schaibley2016,Vitale2018, Liu2019}.  
A hallmark phenomenon in valley materials is the valley Hall effect, where an in-plane electric field drives charge carriers from different valleys to deflect in opposite directions due to valley-contrasting Berry curvature, thereby generating a transverse valley Hall current \cite{Xiao2007,Vitale2018}.
The conventional valley Hall effect is a linear response to the electric field and requires the breaking of crystalline spatial inversion symmetry,
which excludes a vast number of centrosymmetric materials from valleytronic applications \cite{Xiao2012,Wu2013,Mak2014,Lee2016,Barre2019}.
Recent explorations of nonlinear transport phenomena have opened pathways to overcome this fundamental symmetry constraint \cite{Gao2014, Sodemann2015, Du2018, Liu2021, Wang2021, Wang2023, Du2021}.   
The nonlinear valley Hall (NVH) effect has been recently proposed \cite{Das2024, Zhou2025}. The effect is characterized by a quadratic dependence of the transverse valley current on applied electric field. That is, the NVH current, $J^{\text{NVH}}_{a} = J^{v_1}_{a} - J^{v_2}_{a} \propto  E_{b}E_{c}$, where $J^{v_{1,2}}_{a}$ are anomalous Hall currents from different valleys, $E$ is the electric field, and $a$, $b$, $c$ are the Cartesian coordinates.   
The NVH effect can occur in centrosymmetric systems, such as uniaxially strained graphene monolayer \cite{Das2024} and MoS$_2$ bilayer \cite{Zhou2025}, significantly expanding the range of viable materials for valleytronics.

On the other hand, coupling valleys with magnetic order has also led to intriguing physical properties, such as spin-valley coupling \cite{Li2020, Ma2021}, valley splitting \cite{Qi2015, Tong2016, Li2020, Zhou2024}, 
and anomalous valley Hall effect \cite{Chen2024}. 
Accordingly, hundreds of magnetic valley materials have been designed in various magnets \cite{Tong2016, Li2020, Chu2021, Tan2023, Li2024}.
In contrast, previous studies on the NVH effect have been limited to nonmagnetic materials, where spin isn't involved in such valley Hall current.
To achieve multiple controls of the NVH effect and design energy-efficient spin-valley electronic devices, it is highly desirable to incorporate spin degrees of freedom into the NVH effect.
Moreover, although the NVH effect is not constrained by the inversion symmetry, it remains subject to the threefold rotational symmetry. 
In previous studies, an uniaxial strain has been introduced into graphene monolayer \cite{Das2024} and MoS$_2$ bilayer \cite{Zhou2025}, which breaks the  rotational symmetry and activates the NVH effect.
However, applied strain, as an external modulation approach,  possesses limitations, e.g. low strain transfer efficiency and nonuniform strain distribution.
These issues restrict the precision and repeatability of applied strain \cite{Yang2021,Pandey2023}.

To overcome the shortcomings arising from the lack of spin degree of freedom and the requirement for external strain in previous studies on the NVH effect, developing a new mechanism that can effectively introduce spin and offer enhanced controllability is of great importance. 
In this work, we design a magnetically controllable NVH effect in centrosymmetric ferromagnets, by combining the tight-binding model and first-principles calculations.
The introduction of the ferromagnetic order offers a dual functionality, i.e. it not only breaks time-reversal symmetry and correspondingly introduce spin polarization, but also probably breaks the crystalline rotational symmetry with the help of the spin-orbit coupling.
To demonstrate these advantages, we first calculate the NVH conductivities in a simple $d$-orbital tight-binding model. 
The numerical calculations reveal that nonvanishing NVH conductivities indeed emerge in pristine centrosymmetric ferromagnet without applied strain, and the conductivities are contributed by single spin, leading to spin-polarized NVH effect.
The magnitude and sign of the NVH conductivity, together with the direction of spin polarization, depend on the magnetization orientation, which generates rich valley Hall currents with different combinations of current direction and spin polarization.
Furthermore, taking a ferromagnetic VSi$_2$N$_4$ bilayer as a representative material, first-principles calculations show the spin-polarized NVH conductivities are considerable in the bilayer, and they are mainly attributed to small band splittings in the neighborhood of $K_\pm$ valleys. 
The calculated NVH conductivities can be also modulated by the magnetization direction, which supports the model calculations.
These findings establish that centrosymmetric ferromagnets provide powerful platforms for magnetically controllable NVH effect, enabling the generation of spin-polarized valley Hall currents and cooperative manipulation of multiple electronic degrees of freedom, thereby extending valleytronics to centrosymmetric magnetic systems and opening a promising avenue for designing advanced spintronic and valleytronic devices.

{\color{blue}\textit{Tight-binding model and its electronic band structure.}} --
To explore possible magnetically controllable NVH effect, we first construct a spinful three-orbital tight-binding model of a centrosymmetric ferromagnet with a space group of $P\bar{3}m1$ (No.~164), using the MagneticTB package \cite{Zhang2022}. 
The space group includes the spatial inversion symmetry $\mathcal{P}$, threefold rotational symmetry  $C_{3z}$, mirror symmetry $M_{x}$ and twofold rotational symmetry $C_{2x}$, which will play roles in determining nonzero NVH conductivity.
It can be found in many two-dimensional magnetic valley materials with hexagonal lattices, e.g. VSi$_2$N$_4$ bilayer \cite{Liang2023} that will be studied later by first-principles calculations. 
The model considers three $d$ orbitals, i.e.  $d_{z^{2}} ,d_{xz}, d_{yz}$, and incorporates on-site atomic spin-orbit couplings and ferromagnetic exchange interactions with tunable magnetization direction.
The details on the tight-binding model and associate parameters are provided in the Supplemental Material (SM hereafter).


\begin{figure}[htb]
\includegraphics[width=8.5 cm]{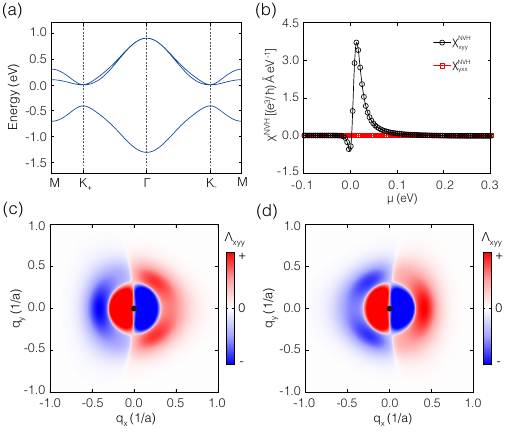}
\caption{ 
Electronic band structure and NVH conductivities of the tight-binding model of a centrosymmetric ferromagnet, with the magnetization along the $x$-axis.
(a) Electronic bands with dominating spin components being along the $x$-axis.
(b) Evolutions of $\chi^{\text{NVH}}_{xyy}$ and $\chi^{\text{NVH}}_{yxx}$ as functions of the chemical potential $\mu$, where the energy window is around the conduction band edge. 
In (a) and (b), the conduction band minima are set to zero energy. 
(c) and (d) Momentum-resolved $\Lambda_{xyy}$ in small regions centered at $K_{+}$ and $K_{-}$, respectively, where $\mu=15$ meV that is near the positive peak of $\chi^{\text{NVH}}_{xyy}$ in (b).  
The locations of $K_\pm$ points are denoted by black dots.
$q_x$ and $q_y$ are wave vectors with respect to $K_\pm$ points.
}

\label{fig01}
\end{figure}

Fig. \ref{fig01}(a) shows a representative band structure of the tight-binding model, with the magnetization along the $x$-axis. Within the energy window of the figure, there are three spin-polarized bands with dominating spin components along the $x$-axis, while the other three bands with opposite spin are out of the energy range owing to strong exchange interactions, as illustrated by spin projections in SM.
According to orbital projections in SM, it is found that the lower band is mainly contributed by $d_{z^2}$ orbital, and two higher bands are composed by $d_{xz}$ and $d_{yz}$ orbitals.
Assuming electrons occupy the lower band, these two groups of bands are referred to as the valence and conduction bands, respectively.
Both the valence and conduction bands reach their local extrema at $K_\pm$ points of the Brillouin zone, indicating $K_\pm$ points correspond to two inequivalent energy valleys.
The band is degenerate between two valleys, which is protected by the $\mathcal{P}$ symmetry that ensures the energy degeneracy at opposite crystal momentum points.
It is different from the $\mathcal{P}$-symmetry-breaking nonmagnetic honeycomb lattice, where the valley degeneracy is protected by the time-reversal symmetry $\mathcal{T}$ \cite{Xiao2012}.  
Moreover, two conduction bands at $K_{\pm}$ valleys have small splittings arising from a weak spin-orbit coupling.
Besides, given that the $C_3$ symmetry excludes nonlinear (valley) Hall effects \cite{Liu2021,Das2024,Zhou2025}, the rotational symmetry with respect to $K_\pm$ points have also been checked.
It is found that the energy contour is no longer $C_3$-symmetric, when in-plane components of the magnetization are present. 
The $C_3$ symmetry breaking allows for nonvanishing NVH conductivities.

{\color{blue}\textit{NVH conductivities based on the tight-binding model.}} --
We then study the NVH effect in centrosymmetric ferromagnet, where the intrinsic band contribution arising from the Berry connection polarizability (BCP) is taken into account.
The intrinsic contribution can be readily distinguished from other extrinsic counterparts by whether it depends on scatterings.
The intrinsic NVH conductivity tensor element, $\chi^{\text{NVH}}_{abc}$, is formulated as \cite{Das2024}
\begin{equation}
\chi^{\text{NVH}}_{abc} = e^{3}\int_{\text{BZ}} \tau \frac{d\bm{k}}{(2\pi)^{2}} \Lambda_{abc}(\bm{k}),  \\
\label{eq1}
\end{equation}
with \begin{align}
\Lambda_{abc}(\bm{k}) &= \sum_{n} \lambda_{abc}^{n}(\bm{k}) \frac{\partial f(\varepsilon_{n\bm{k}})}{\partial \varepsilon_{n\bm{k}}}, \\
\lambda^{n}_{abc}(\bm{k}) &=  v^{n}_{a}G^{n}_{bc}(\bm{k})-v^{n}_{b}G^{n}_{ac}(\bm{k}),
\end{align}
 \begin{eqnarray}
G_{ab}^{n}(\bm{k}) &=& 2 \text{Re} \sum_{m\ne n} \frac{A^{nm}_{a}(\bm{k})A^{mn}_{b}(\bm{k})}{\varepsilon_{n\bm{k}}-\varepsilon_{m\bm{k}}},\\
\label{eq5}
A^{nm}_{a}(\bm{k}) &=& \frac{i\hbar\langle u_{n\bm{k}} | \hat{v}_{a} | u_{m\bm{k}} \rangle}{{\varepsilon_{m\bm{k}}-\varepsilon_{n\bm{k}}}},
\end{eqnarray} 
where $\tau = \pm 1$ denote $K_\pm$ valleys, respectively, and distinguish contributions from the two valleys in Eq. \ref{eq1}. $\Lambda_{abc}$, $\lambda^{n}_{abc}$ and $G^{n}_{ab}$ are the BCP dipole, band-resolved BCP dipole and BCP, respectively.  $A^{nm}_{a}$ is the interband Berry connection, and defined in terms of the periodic part of Bloch electronic state $|u_{n\bm{k}}\rangle$ and band energy $\varepsilon_{n\bm{k}}$, with $\bm{k}$ being wave vector and $n$, $m$ being band indices.
$f(\varepsilon_{n\bm{k}})$ is the equilibrium Fermi-Dirac distribution, and $v^{n}_{a}=\langle u_{n\bm{k}} | \hat{v}_{a} | u_{n\bm{k}} \rangle$ is the  expectation of the velocity operator $\hat{v}_{a}$.
The subscript $a$ denote the direction of NVH current, while $b$ and $c$ denote directions of electric fields.   
Considering the experimental setup of two-dimensional Hall transport, the above directions are set to in-plane ones, with $b=c$. That is, only $\chi_{xyy}^{\text{NVH}}$ and $\chi_{yxx}^{\text{NVH}}$ are considered in our calculations. 
Based on calculated $\chi_{abb}^{\text{NVH}}$, the NVH conductivity can be then obtained as  $J^{\text{NVH}}_{a} = J^{K_{+}}_{a} - J^{K_{-}}_{a} = \chi^{\text{NVH}}_{abb} E_{b}^2$.
 
Fig. \ref{fig01}(b) shows calculated NVH conductivity tensor elements, $\chi^{\text{NVH}}_{xyy}$ and $\chi^{\text{NVH}}_{yxx}$, as functions of chemical potential $\mu$, where the magnetization is along the $x$-axis.  
The NVH conductivity is given in the energy range near the conduction band edges, because of the NVH conductivity in the range is much remarkable compared with that in the energy range of the valence band.
It is seen that while $\chi^{\text{NVH}}_{yxx}$ is always zero, $\chi^{\text{NVH}}_{xyy}$ is nonvanishing and varies significantly with $\mu$. 
As $\mu$ increases, $\chi^{\text{NVH}}_{xyy}$ first exhibits a small negative peak, then reverses sign and sharply amplifies into a large positive peak before asymptotically vanishing. The sign reversal is a characteristic signature of nonlinear (valley) Hall effect generated by gapped Dirac fermions \cite{Du2018, Zhou2025}. 
Besides, the peak energies of the NVH response are close to those associated with small band gaps between the conduction bands at $K_\pm$ valleys. 
The sizable NVH response from small-gap regions is due to the scaling relations $A_{\alpha}^{nm} \propto (\varepsilon_{nk}-\varepsilon_{mk})^{-1}$ in Eq. \ref{eq5} and resultant $G_{\alpha\beta}^{n} \propto (\varepsilon_{nk}-\varepsilon_{mk})^{-3}$ \cite{Du2018, Zhou2025}.

To elucidate contributions from $K_\pm$ valleys to the NVH conductivities, momentum-resolved BCP dipole, $\Lambda_{xyy}(\bm{k})$, in the neighborhood of the two valleys are present in Figs. \ref{fig01}(c) and \ref{fig01}(d), respectively, and the plots of $\Lambda_{yxx}(\bm{k})$ are given in SM.
It is seen that $\Lambda(\bm{k})$ is pronounced near $K_\pm$ valleys and gradually decreases when moving away from the two valleys, which confirms the dominating contributions from $K_\pm$ valleys.
According to the momentum-resolved BCP dipole, finite $\chi^{\text{NVH}}_{xyy}$ and vanishing $\chi^{\text{NVH}}_{yxx}$ can be also explained from the view of the symmetry.
Considering the magnetization along the $x$-axis, the $C_{2x}$ and $M_x$ symmetries are kept in the ferromagnet.
With respect to the line $q_y = 0$, $\Lambda_{xyy}(\bm{k})$ is found to be symmetric, which is protected by the $C_{2x}$ symmetry \cite{Zhou2025}.
The detailed symmetry analyses are provided in SM.
The symmetric $\Lambda_{xyy}(\bm{k})$ allows for nonzero nonlinear Hall response at each valley.
Furthermore, the $\mathcal{P}$ ($M_x$) symmetry ensures opposite $\Lambda_{xyy}(\bm{k})$ at $\mathcal{P}$-paired ($M_x$-paired) wave vectors from two valleys, leading to opposite nonlinear Hall response and consequent nonvanishing $\chi^{\text{NVH}}_{xyy}$.
In contrast, $\Lambda_{yxx}$ is antisymmetric with respect to $q_y=0$ at each valley.
As a result, the nonlinear Hall response vanishes at each valley, and $\chi^{\text{NVH}}_{yxx}$ is zero.

\begin{figure}[htb]
\includegraphics[width=8.5 cm]{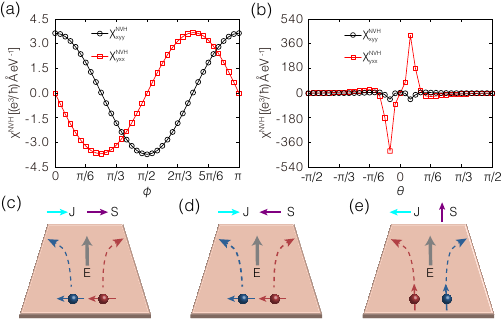}
\caption{ 
Magnetically controllable NVH conductivities.
(a) Evolutions of $\chi^{\text{NVH}}_{xyy}$ and $\chi^{\text{NVH}}_{yxx}$ with the angles $\phi$ and $\theta$ at $\mu=15$ meV, respectively.
$\theta$ and $\phi$ are set to zero in (a) and (b), respectively. 
(c)-(e) Schematic depictions of the NVH effect with the magnetization along the $x$, $-x$, and $y$ directions, respectively.  
Red and blue balls stand for carriers from $K_+$ and $K_-$ valleys, respectively, with arrows representing spin directions. 
Cyan and purple arrows denote directions of the NVH current and the spin polarization, respectively. 
}
\label{fig02}
\end{figure}
We further investigate the effect of the magnetization direction on NVH conductivities. 
The magnetization direction is defined by the azimuthal angle, $\phi$, and the canting angle, $\theta$, with $\theta=0$ denoting the in-plane magnetization.
Fig. \ref{fig02}(a) shows evolutions of NVH conductivity tensor elements with the angle $\phi$ on the equator of the unit sphere representing the magnetization direction ($\theta=0$).
It is found that both $\chi^{\text{NVH}}_{xyy}$ and $\chi^{\text{NVH}}_{yxx}$ exhibit a $\pi$ periodicity, and they approximately follow cosine and sine functions, respectively.
Similar to the magnetization along the $x$-axis  ($\phi=0$), the magnetization along the $y$-axis ($\phi = \pi/2$) also has nonzero $\chi^{\text{NVH}}_{xyy}$ and vanishing $\chi^{\text{NVH}}_{yxx}$.
This is protected by $M_x\mathcal{T}$ and $\mathcal{P}$ symmetries,
where $M_x\mathcal{T}$ plays the role of $C_{2x}$ in the magnetization along the $x$-axis, respectively (see SM).
When the magnetization moves away from $x$- and $y$-axes, it is seen that both $\chi^{\text{NVH}}_{xyy}$ and $\chi^{\text{NVH}}_{yxx}$ are nonvanishing, owing to the breakings of $C_{2x}$- and $M_x$-related symmetries.

Fig. \ref{fig02}(b) presents variations of NVH conductivity tensor elements with the angle $\theta$, in a meridional plane of the unit sphere of the magnetization direction with $\phi=0$. 
Compared with those in Fig. \ref{fig02}(a), $\chi^{\text{NVH}}_{xyy}$ and $\chi^{\text{NVH}}_{yxx}$ in the figure exhibit more complicated angular dependence behaviors.
It is noted that with the magnetization along the $z$ direction, both $\chi^{\text{NVH}}_{xyy}$ and $\chi^{\text{NVH}}_{yxx}$ are computed to be zero, similar to the cases of pristine graphene monolayer and MoS$_2$ bilayer without applied strain \cite{Das2024, Zhou2025}.
This is due to the presence of the $C_3$ symmetry.
When moving away from the $z$ axis, the NVH responses become nonvanishing.
Therefore, the NVH effect in centrosymmetric ferromagnet requires  in-plane components of the magnetization. 
Moreover, $\chi^{\text{NVH}}_{xyy}$ is even with respect to $\theta=0$, while $\chi^{\text{NVH}}_{yxx}$ is odd, which is protected by $M_x$ and $\mathcal{P}$ symmetry operations (see SM).
Besides, $\chi^{\text{NVH}}_{xyy}$ and $\chi^{\text{NVH}}_{yxx}$ have peaks at the angle, $\theta=11^\circ$.

These above findings are intriguing. 
Firstly, the introduction of in-plane magnetization components breaks the $C_3$ symmetry and enables nonvanishing NVH conductivities. 
This is an intrinsic method based on the ferromagnetic order that obviates the need for external strain. 
Secondly, NVH conductivities exhibit  strong dependencies on the magnetization direction. 
By rotating the magnetization, one can not only effectively modulate the magnitude of the nonlinear response, but also realize its sign reversal, thereby providing an exotic magnetically controllable NVH transport.

Thirdly, the NVH effect is spin-polarized with multiple modes.
Given that the spin direction is determined by the magnetization direction, its variation exhibits a $2\pi$ periodicity upon the in-plane magnetization rotation, in contrast to the $\pi$ periodicity of NVH conductivities in Fig. \ref{fig02}(a).
The different periodicities result in various combinations of NVH currrent and spin polarization.
We take $\chi^{\text{NVH}}_{xyy}$ for example, where the NVH current flows along the $x$ direction under applied electric field along the $y$ direction, as illustrated in Figs. \ref{fig02}(c-e).
When reversing the magnetization direction, the NVH current is unchanged but the spin polarization becomes opposite.
The spin-polarized currents with opposite spins can thus be obtained [see Figs. \ref{fig02}(c) and \ref{fig02}(d)]. 
When modulating the magnetization from the $x$-axis to $y$-axis, the NVH current reverses direction and the spin rotates by $\pi/2$.
As a result, spin-polarized currents with spin both parallel and perpendicular to the current can be generated [see Figs. \ref{fig02}(c) and \ref{fig02}(e)]. 
These rich spin-polarized NVH currents are expected to be utilized for information encoding and manipulation based on valley and spin degrees of freedom.

{\color{blue}\textit{Application to VSi$_2$N$_4$ bilayer.}} --By first-principle based calculations, we then study material realizations of magnetically controllable NVH effect, where a bilayer of ferromagnetic VSi$_2$N$_4$ is taken for example.
The first-principles calculation details can be found in SM.
In previous work, the most stable stacking of the bilayer is the AA$^\prime$ one, and interlayer ferromagnetic and antiferromagnetic couplings are nearly degenerate in energy and can be easily switched by applied magnetic field \cite{Liang2023}.
Accordingly, we consider the AA$^\prime$-stacked VSi$_2$N$_4$ bilayer with both intralayer and interlayer ferromagnetic order.
The bilayer has the space group, $P\bar{3}$m1, which agrees with our tight-binding model of centrosymmetric ferromagnet.

\begin{figure}[htb]
\includegraphics[width=8.5 cm]{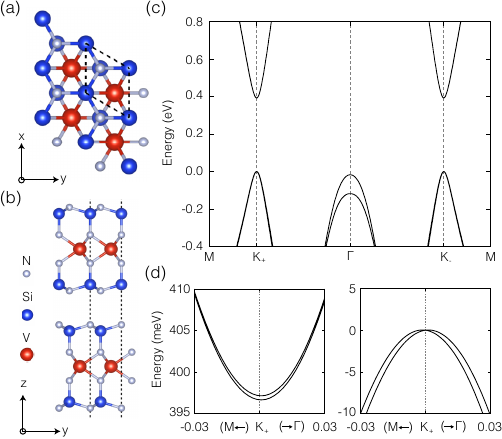}
\caption{ Crystal and electronic band structures of the  AA$^\prime$-stacked VSi$_2$N$_4$ bilayer.
(a) Top view and (b) side view of the bilayer. 
Red, blue and white balls stand for V, Si and N atoms, respectively.
(c) Band structure of the bilayer, with the magnetization along the $x$-axis. 
(d) Enlarged band structures around the conduction and valence band edges. 
In (c) and (d), the valence band maxima are set to zero energy.
}
\label{fig03}
\end{figure}

Figs. \ref{fig03}(a) and \ref{fig03}(b) shows top and side views of the AA$^\prime$-stacked VSi$_2$N$_4$ bilayer.
Each monolayer in the bilayer structure exhibits a two-dimensional hexagonal lattice and comprises seven atomic layers with a stacking sequence of N–Si–N–V–N–Si–N.
The VSi$_2$N$_4$ monolayer can be regarded as an VN$_2$ layer sandwiched between two SiN layers, with the VN$_2$ layer being isostructural to transition metal dichalcogenides in the 2H phase \cite{Xiao2012}.
In the AA$^\prime$-stacked VSi$_2$N$_4$ bilayer, two monolayers are aligned with a 180$^\circ$ rotation relative to each other.
 As viewed from above, V atoms of two monolayers are directly superposed, while N atoms of the upper (lower) VN$_2$ layer are centered on hexagonal rings of the lower (upper) VN$_2$ layer. 
For the bilayer, the in-plane lattice constant is computed to be 2.89 \AA. The monolayer thickness is 6.87 \AA, and the van der Waals gap between two monolayers is 2.80 \AA. 
The total magnetic moment in a unit cell is 2 $\mu_B$, with each V atom contributing 1.19 $\mu_B$ and small contributions from other atoms.
These results are similar to previous work \cite{Liang2023}.
Moreover, it is noted that the interlayer ferromagnetic coupling in the bilayer preserves the $\mathcal{P}$ symmetry for arbitrary magnetization direction.
The symmetry results in the full suppression of the linear valley Hall response, while the NVH effect is symmetry-allowed, making the bilayer an ideal platform for investigating the nonlinear response.

Fig. \ref{fig03}(c) demonstrates a representative band structure of the VSi$_2$N$_4$ bilayer, with the magnetization is along the $x$-axis.
Within the energy range considered here, there are two conduction bands and two valence bands.
Similarly to the tight-binding band structure in Fig. \ref{fig01}(b), the valence band maxima and conduction band minima appear at inequivalent $K_\pm$ valleys.
Besides, small band gaps are found between two conduction bands and between two valence bands in the vicinity of $K_\pm$ valleys, as illustrated in Fig. \ref{fig03}(d).
The small band gaps arise from weak interaction between two VSi$_2$N$_4$ monolayers, and they are anticipated to result in considerable nonlinear responses, as mentioned above.
\begin{figure}[htb]
\includegraphics[width=8.5 cm]{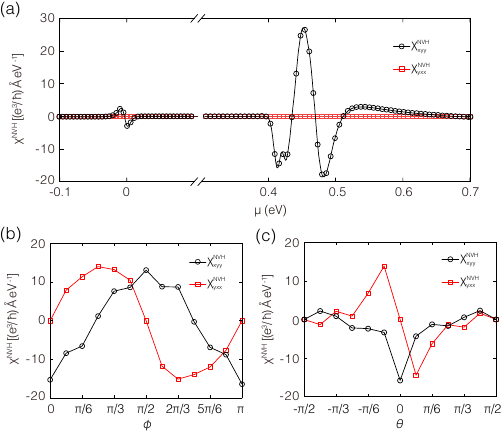}
\caption{ NVH conductivities in the VSi$_2$N$_4$ bilayer.
(a) Evolutions of $\chi^{\text{NVH}}_{xyy}$ and $\chi^{\text{NVH}}_{yxx}$ as functions of the chemical potential $\mu$, with the valence band maxima set to zero energy.
(b) Evolutions of $\chi^{\text{NVH}}_{xyy}$ and $\chi^{\text{NVH}}_{yxx}$ as functions of the angle $\phi$, with $\theta=0$.
(c) Evolutions of $\chi^{\text{NVH}}_{xyy}$ and $\chi^{\text{NVH}}_{yxx}$ as functions of the angle $\theta$, with $\phi=0$.
$\mu$ is chosen to be 415 meV in (b) and (c).
}
\label{fig04}
\end{figure}

The NVH conductivities are then calculated for the VSi$_2$N$_4$ bilayer, based on the tight-binding Hamiltonian generated by the Wannier90 package with first-principles input \cite{Pizzi2020}.
Fig. \ref{fig04}(a) displays NVH conductivity tensor elements as functions of the chemical potential $\mu$, where the magnetization is along the $x$-axis.
It is seen that while $\chi^{\text{NVH}}_{yxx}$ is always zero, $\chi^{\text{NVH}}_{xyy}$ exhibits considerable, dipole-like peaks with sign reversals near both the conduction and valence band edges.
The peak values of the conduction and valence bands are of the order of $10^0$ and $10^1$ $(e^{3}/\hbar)$\AA\ eV$^{-1}$, respectively, which surpass values of NVH conductivities in strained graphene \cite{Das2024} and nonlinear Hall conductivities of CuMnAs film estimated by experimentally observed second-order magnetoresistance effect \cite{Godinho2018}.   
The sizable nonlinear response is attributed to small band splittings between two conduction (valence) bands, and expected to be measured in experiment. 

Fig. \ref{fig04}(b) and \ref{fig04}(c) further demonstrate evolutions of $\chi^{\text{NVH}}_{xyy}$ and $\chi^{\text{NVH}}_{yxx}$ in the VSi$_2$N$_4$ bilayer with in-plane and out-plane rotations of the magnetization, respectively. 
For the in-plane magnetization rotation, $\chi^{\text{NVH}}_{xyy}$ and $\chi^{\text{NVH}}_{yxx}$ behave like the functions of $-$cos2$\phi$ and sin2$\phi$, respectively.
When the magnetization rotates out-of-plane,  $\chi^{\text{NVH}}_{xyy}$ and $\chi^{\text{NVH}}_{yxx}$  display more complex angular dependencies.
They are even and odd with respect to $\theta=0$, respectively.
The magnetically controllable NVH conductivities in the VSi$_2$N$_4$ bilayer support the findings based on the tight-binding model. 

{\color{blue}\textit{Discussion and summary.}} --
It is noted that the VSi$_2$N$_4$ bilayer studied above is only one instance that exhibits magnetically controllable NVH effect. This effect is expected to apply to a broader class of centrosymmetric magnetic valley materials, across hexagonal, square, or rectangular lattices.
Even when focusing on a hexagonal lattice like the VSi$_2$N$_4$ bilayer, there are also a number of choices of symmetry groups, such as $C_{6h}$-, $D_{3h}$- and $D_{6h}$-derived space groups.  
Regarding specific candidate materials, systems such as the Cr$_2$Ge$_2$Te$_6$ monolayer (space group $P\bar{3}1m$) \cite{Wang2019}, 
EuSi$_2$ layers ($P6/mmm$) \cite{Wang2025} and transition metal trihalide monolayers ($P\bar{3}m1$) \cite{tomar2019} all possess both $\mathcal{P}$ and $C_3$ symmetries, and they are likely to generate magnetically controllable NVH conductivities and worthy of further investigation.   
Furthermore, given that hundreds of magnetic valley monolayers have been proposed \cite{Tong2016, Li2020, Chu2021, Tan2023, Li2024}, their homobilayers with the $\mathcal{P}$ symmetry can also be constructed via a 180$^\circ$ twist between two monolayers, and they support NVH effect rather than linear valley Hall effect.

For centrosymmetric ferromagnets, the presence of the $\mathcal{P}$ symmetry precludes the influence of the linear valley Hall effect on the observation of the NVH effect.
However, linear anomalous Hall (LAH) effect might persist. 
The NVH and LAH effects have distinct characteristics and they are  experimentally distinguishable.
Firstly, since these two effects are nonlinear and linear responses to applied electric field, an a.c. current can be utilized to generate Hall voltages at the fundamental and second-harmonic frequencies, respectively, which can be readily distinguished using a lock-in amplifier \cite{Sodemann2015}.
Secondly, these two effects have different dependences on the magnetization direction. 
While the LAH  conductivity typically scales with the out-of-plane component of the magnetization \cite{Nagaosa2010}, 
nonvanishing NVH conductivities require the presence of in-plane magnetization components. This distinction also enables the differentiation of the NVH effect from the LAH effect. 

To conclude, a magnetically controllable NVH effect is studied by the tight-binding and first-principles calculations.
The effect generates various spin-polarized valley Hall currents, with its magnitude, sign and spin polarization highly tunable by the magnetization orientation, and it is manifested in a ferromagnetic VSi$_2$N$_4$ bilayer. 
The NVH transport is expected to be observable via nonlocal resistance measurement \cite{Sui2015, Shimazaki2015},
and utilized for designing energy-efficient valleytronic and spintronic devices based on manipulations of valley and spin degrees of freedom.

{\color{blue}\textit{Acknowledgments}} --  
We are supported by the National Natural Science Foundation of China (Nos. 12374044, 12004186, 11904173). 

%


\providecommand{\noopsort}[1]{}

\end{document}